\documentclass[twocolumn]{aastex62}
\usepackage{apjfonts,amsmath,verbatim}
\usepackage[normalem]{ulem}

\usepackage{soul}
\shorttitle{The Role of Strong Gravity and the Nuclear EoS on NS CE Accretion}
\shortauthors{Holgado et al.}

\graphicspath{{./figures/}}

\begin{document}

\title{THE ROLE OF STRONG GRAVITY AND THE NUCLEAR EQUATION OF STATE \\ ON NEUTRON-STAR COMMON-ENVELOPE ACCRETION}

\correspondingauthor{A.~Miguel Holgado}
\email{mholgado@andrew.cmu.edu}

\author[0000-0003-4143-8132]{A.~Miguel Holgado}
\affil{Department of Physics and McWilliams Center for Cosmology, Carnegie Mellon University, Pittsburgh, PA, 15213, USA
}
\affil{Department of Astronomy and National Center for Supercomputing Applications, University of Illinois at Urbana-Champaign, Urbana IL, 61801, USA
}
\author[0000-0002-0066-9471]{Hector O.~Silva}
\affiliation{Max-Planck-Institut f\"ur Gravitationsphysik (Albert-Einstein-Institut),
Am M\"uhlenberg 1, D-14476 Potsdam, Germany}
\affil{Department of Physics and Illinois Center for Advanced Studies of the Universe, University of Illinois at Urbana-Champaign, Urbana IL, 61801, USA
}
\author[0000-0002-5294-0630]{Paul M.~Ricker}
\affil{Department of Astronomy and National Center for Supercomputing Applications, University of Illinois at Urbana-Champaign, Urbana IL, 61801, USA
}
\affil{Department of Physics and Illinois Center for Advanced Studies of the Universe, University of Illinois at Urbana-Champaign, Urbana IL, 61801, USA
}
\author[0000-0001-6147-1736]{Nicol\'{a}s Yunes}
\affil{Department of Physics and Illinois Center for Advanced Studies of the Universe, University of Illinois at Urbana-Champaign, Urbana IL, 61801, USA
}



\begin{abstract}
Common-envelope evolution is important in the formation of neutron star binaries within the isolated binary formation channel.
As a neutron star inspirals within the envelope of a primary massive star, it accretes and spins up. 
Because neutron stars are in the strong-gravity regime, they have a substantial relativistic mass deficit, i.e., their gravitational mass is less than their baryonic mass. 
This effect causes some fraction of the accreted baryonic mass to convert into neutron star binding energy. 
The relativistic mass deficit also depends on the nuclear equation of state, since more compact neutron stars will have larger binding energies. 
We model the mass growth and spin-up of neutron stars inspiraling within common-envelope environments and quantify how different initial binary conditions and hadronic equations of state affect the post-common-envelope neutron star's mass and spin. 
From these models, we find that neutron star mass growth is suppressed by $\approx 15-30\%$. 
We also find that for a given amount of accreted baryonic mass, more compact neutron stars will spin-up faster while gaining less gravitational mass, and vice versa. 
This work demonstrates that a neutron star's strong gravity and nuclear microphysics plays a role in neutron-star-common-envelope evolution, in addition to the macroscopic astrophysics of the envelope. 
Strong gravity and the nuclear equation of state may thus affect both the population properties of neutron star binaries and the cosmic double neutron star merger rate. 
\end{abstract}

\keywords{neutron stars --- common envelope evolution --- accretion --- nuclear physics --- compact objects -- compact binary stars -- interacting binary stars}

\section{Introduction} \label{sec:intro}
Neutron stars (NSs) as well as double neutron star (DNS) systems are versatile laboratories for multiple disciplines, including (but not limited to) astrophysics, nuclear physics, and gravitational physics. 
Our knowledge of DNS population properties as well as the nuclear equation of state (EoS) has greatly improved as we are entering a data-rich era for NS observations. 
The LIGO-Virgo Collaboration (LVC) has observed GWs from NS mergers, providing constraints on the NS tidal deformability \citep{ligo_scientific_collaboration_and_virgo_collaboration_gw170817:_2017,ligo_scientific_collaboration_and_virgo_collaboration_properties_2019} and new insights on the DNS mass distribution \citep{abbott_gw190425_2020}.
NICER X-ray timing observations of pulsars have provided the first constraints on the NS compactness \citep{miller_psr_2019,riley_nicer_2019}.  
Radio pulsar timing has revealed the most massive NS to date from the Green Bank Telescope \citep{cromartie_relativistic_2020} and has also revealed a DNS with the lowest asymmetric mass ratio of $0.78 \pm 0.03$ observed to date from the Arecibo Observatory \citep{ferdman_asymmetric_2020}.
\par
A DNS that forms in isolation must survive two supernova explosions and one or more common-envelope (CE) phases \citep[e.g.,][]{andrews_evolutionary_2015,tauris_formation_2017,andrews_double_2019}. 
In the context of CE evolution, NSs have been treated as point masses that accrete some fraction of their pre-CE mass, similar to white dwarfs and black holes \citep[e.g.,][]{belczynski_population_2002,belczynski_comprehensive_2002,voss_galactic_2003,dewi_double-core_2006,oslowski_population_2011,dominik_double_2012,belczynski_origin_2018,chruslinska_double_2018,giacobbo_progenitors_2018,vigna-gomez_common_2020,kruckow_masses_2020}. 
The NS's strong gravity and nuclear EoS, however, result in a relativistic mass deficit, where the gravitational mass is significantly less than the total baryonic mass. 
This binding-energy effect has been previously studied in the context of NS accretion in low-mass X-ray binaries \citep{alecian_effect_2004,lavagetto_role_2005,bagchi_role_2011}.
\par
Early theoretical studies of NS mass growth during CE evolution predicted that accretion would be substantial enough to cause NSs to collapse into black holes \citep[e.g.,][]{chevalier_neutron_1993,brown_neutron_1995,fryer_dynamics_1996,armitage_black_2000,brown_hypercritical_2000}.
Global 3D hydrodynamic CE simulations, however, have found typical accretion rates to be less than the Hoyle-Lyttleton (HL) rate \citep[e.g.,][]{ricker_amr_2012}. 
Moreover, \cite{macleod_asymmetric_2015} have found from local 3D wind-tunnel simulations that envelope density gradients may substantially suppress the accretion rate \citep{macleod_asymmetric_2015}.
These results imply that NSs are much more likely to survive the CE phase \citep[e.g.,][]{macleod_accretion-fed_2015,holgado_gravitational_2018} instead of collapsing into black holes.
\par
Further wind-tunnel studies have provided more insights into how the local density gradient and flow properties are correlated, where such correlations occur, and to what extent such correlations hold \citep{macleod_common_2017,de_common_2020,everson_common_2020}.
General-relativistic 2D wind-tunnel simulations with a relativistic plasma have also been carried out to characterize accretion and drag on compact-object scales \citep{cruz-osorio_common-envelope_2020}. 
Building on these general-relativistic models towards 3D and further capturing the plasma conditions relevant to massive-star interiors is certainly well motivated. 
In addition to these studies of accretion and drag local to the compact object, the global numerical modeling of NS-CE evolution has been steadily progressing with 1D hydrodynamic \citep{fragos_complete_2019} and 3D hydrodynamic models \citep{law-smith_successful_2020}.  
\par
As such numerical models improve in complexity, it may soon be of interest to consider how CE evolution may be sensitive to additional physics, which itself is an open question. 
Given the current observational constraints on the nuclear EoS, we here investigate how a NS's macrosopic properties affects its mass-growth and spin-up during the CE inspiral, and before the primary explodes and forms another NS. 
In addition to focusing on the role of strong gravity and the nuclear EoS, we approximate the pertinent aspects of the accretion and local dynamical friction, which isolates the full complexities of the macroscopic CE physics. 
\section{Methods} \label{sec:methods}
We consider a primary massive star with mass $M_\star$ and radius $R_\star$ orbiting a companion NS with initial mass $M_{\rm NS,0}$ that rotates rigidly with an initial angular frequency $\Omega_0$.
For the system to be in the NS-CE phase, we also initialize the orbit at a separation $a_0$ that is equal to the radius of the primary massive star, $a_0 = R_\star$.
The primary's radius $R_\star$ will depend on its evolutionary stage, where we consider here the base and tip of the red-giant branch (RGB). 
\par
During the CE phase, the inspiral is driven by local dynamical friction, causing the NS to accrete matter and spin-up. 
If enough energy is injected into the CE, it will be ejected, thus leaving a less massive primary star and a spun-up NS at a closer separation; the DNS then would form after the primary goes supernova. 
A second CE, however, may occur before the primary helium star forms the second NS \citep[e.g.,][]{dewi_evolution_2002,ivanova_common_2013,romero-shaw_origin_2020,galaudage_heavy_2021}, though we leave such considerations for future work. 
\subsection{Neutron Star Equation of State, Stellar Structure, Accretion, and Spin-Up}
Even for the highest spinning pulsars observed to date, such NSs can be considered as slowly-rotating objects, 
meaning that rotation can thus be treated as a small perturbation $\epsilon$ to the Tolman-Oppenheimer-Volkoff (TOV) solution for non-rotating NSs \citep{tolman_static_1939,oppenheimer_massive_1939}. 
Here, $\epsilon \equiv \Omega/\Omega_{\rm k}$ is a dimensionless spin parameter, where $\Omega$ is the angular spin frequency of the star, and $\Omega_{\rm k}$ is the Keplerian angular spin frequency $\Omega_{\rm k} = \sqrt{G M_{\rm TOV}/R_{\rm TOV}^3}$, with $M_{\rm TOV}$ and $R_{\rm TOV}$ the mass and radius of our NS if it were not rotating. 
We solve for the structure of slowly rotating NSs to second-order in $\epsilon \ll 1$ using the Hartle-Thorne approximation \citep{hartle_slowly_1967,hartle_slowly_1968} with the same set of 46 hadronic EoSs from \citet[][\autoref{app:eos}]{silva_astrophysical_2020}.
This set of EoSs is simultaneously consistent with the  LIGO-Virgo observations of GW170817 \citep{ligo_scientific_collaboration_and_virgo_collaboration_gw170817:_2017} and the NICER observation of PSR J0030+0451 \citep{miller_constraining_2019,riley_nicer_2019}. 
\par
For a given EoS, and a chosen value of the central density and spin frequency $\Omega$, the second-order in $\epsilon$ solution to the Einstein equations in the Hartle-Thorne approximation allows us to calculate macroscopic properties of the star \citep[e.g.,][]{berti_rotating_2005}. These properties include the spin-corrected mass $M_{\rm NS} = M_{\rm TOV} + \epsilon^2 \delta M$, the spin-corrected equatorial radius  $R_{\rm NS} = R_{\rm TOV} + \epsilon^2 \delta R$, the leading-order-in-spin moment of inertia $I_{\rm NS}$
and the spin-corrected dimensionless gravitochemical potential $\Phi_{\rm NS}$~\citep{alecian_effect_2004}. 
\par
In the context of accreting NSs, the gravitochemical potential can be interpreted as a susceptibility to changes in baryon number or the fraction of baryon mass that gets converted into gravitational mass. 
In the non-rotating limit, $\Phi_{\rm NS}$ simplifies to
\begin{equation} \label{eq:non}
\lim_{\Omega\to 0} \Phi_{\rm NS} = \Phi_{\rm TOV} = \sqrt{1 - 2 {\cal C}_{\rm TOV}} = \sqrt{1 - \frac{2 G M_{\rm TOV}}{c^2 R_{\rm TOV}}} \ ,
\end{equation}
where ${\cal C}_{\rm TOV}$ is the compactness of a given non-rotating NS. 
We will use $\Phi_{\rm NS}$ for our calculations, where we elaborate in \autoref{app:phi} how this is calculated with our EoS catalog. 
\autoref{eq:non} provides a fast approximation for population synthesis or as a sub-grid prescription for global hydrodynamic simulations. 
We later compare in \S\ref{sec:results} how well this approximation compares to using $\Phi_{\rm NS}$, with a more detailed quantification shown in \autoref{app:kl}. 
\par
\begin{figure}
\centering
\includegraphics[width=\columnwidth]{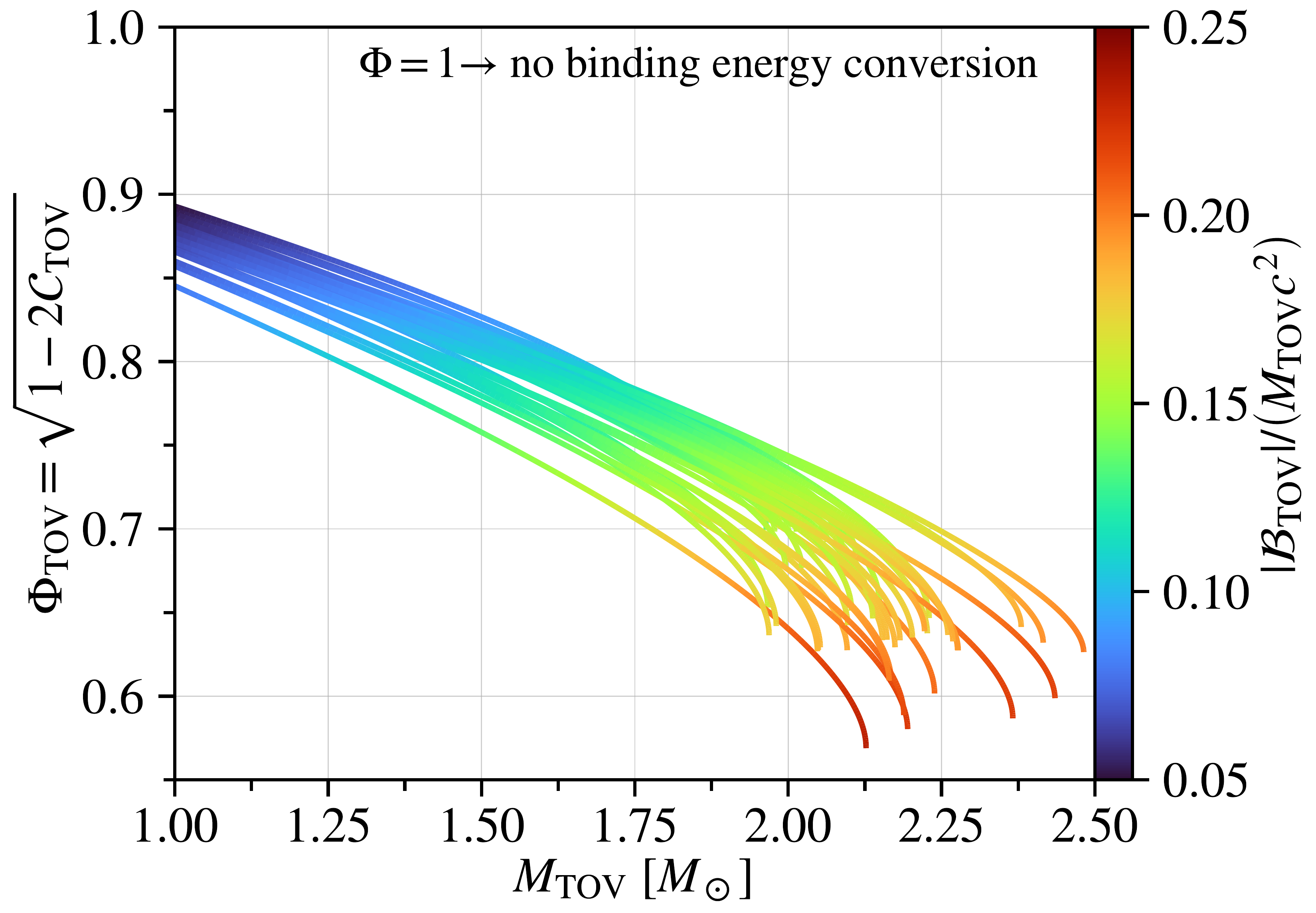}
\caption{\label{fig:phis} {\bf Gravitochemical potential vs.~gravitational mass for non-rotating NSs.} 
Each curve corresponds to a different EoS in our catalog that is consistent with both the latest LIGO-Virgo and NICER constraints and is able to produce a NS with $M_{\rm max}/M_\odot \ge 1.96$.
The color of each curve corresponds to the nondimensional NS binding energy $|{\cal B}_{\rm TOV}|/(M_{\rm TOV}c^2)$. 
For $\Phi = 1$, all of the accreted baryonic mass contributes to the gravitational mass growth. 
NSs with larger gravitational masses and with larger compactness will convert a larger fraction of the accreted baryonic mass into binding energy instead of gravitational mass.
}
\end{figure}

We plot in \autoref{fig:phis} the gravitochemical potential $\Phi_{\rm TOV}$ versus the gravitational mass $M_{\rm TOV}$ for non-rotating NSs, as predicted from our EoS catalog. 
Each curve represents $\Phi_{\rm TOV}$ for a different EoS, with different colors corresponding to different dimensionless NS binding energy $|{\cal B}_{\rm TOV}|/(M_{\rm TOV}c^2)$. 
Observe that the gravitochemical potential decreases as the gravitational mass increases, and also it decreases as the NS compactness increases. 
Since the gravitochemical potential is inversely related to the binding energy, this figure tells us that binding energy conversion is enhanced for higher mass or higher compactness NSs.  
\par
As the NS mass and spin increase during the CE inspiral, we are then able to track the temporal evolution of all of NS macroscopic quantities. 
For example, as the NS accretes, its gravitational mass responds to the baryon mass accretion rate $\dot{M}_{\rm b}$ as well as the angular momentum that the accreted mass carries. The resulting NS gravitational-mass accretion rate is
\begin{equation}  \label{eq:dM1}
\dot{M}_{\rm NS} = \frac{\partial M_{\rm NS}}{\partial M_{\rm b}} \dot{M}_{\rm b} + \frac{\partial M_{\rm NS}}{\partial J_{\rm NS}} \dot{J}_{\rm NS} = \Phi_{\rm NS} \dot{M}_{\rm b} + \frac{\Omega}{c^2} \dot{J}_{\rm NS} \ ,
\end{equation}
where $c$ is the speed of light, and $J_{\rm NS} = I_{\rm NS} \Omega$ is the NS spin angular momentum. Similarly, as the NS accretes, its spin angular momentum will also change, as given by
\begin{equation} \label{eq:dJ}
\dot{J}_{\rm NS} = \dot{I}_{\rm NS} \Omega + I_{\rm NS} \dot{\Omega} = \frac{{\rm d}I_{\rm NS}}{{\rm d}M_{\rm NS}} \dot{M}_{\rm NS} \Omega + I_{\rm NS} \dot{\Omega} \ .
\end{equation}
%
%
%
\par
%
%
With this at hand, we can now solve for the temporal evolution of the angular frequency and the gravitational mass. We assume the NS accretes from a Keplerian accretion disk, where matter captured within the NS accretion radius carries angular momentum and spirals several orders of magnitude down to the scale of several NS radii. 
Approximating the total torque as the accretion torque, $\dot{J}_{\rm NS} \approx \dot{M}_{\rm NS} \sqrt{G M_{\rm NS} R_{\rm NS}}$ \citep[e.g.,][]{brown_hypercritical_2000}, in Eqs.~\ref{eq:dM1} and~\ref{eq:dJ}, and for now ignoring other external torques on the NS, we then find
\begin{equation} \label{eq:dM}
\dot{M}_{\rm NS} = \frac{\Phi_{\rm NS} \dot{M}_{\rm b}}{1 - \sqrt{G M_{\rm NS} R_{\rm NS}}\, ({\Omega}/{c^2})} \ ,
\end{equation}
and 
\begin{equation} \label{eq:dOm}
\dot{\Omega} = \frac{\Phi_{\rm NS} \dot{M}_{\rm b}}{I_{\rm NS}}
\frac{\sqrt{G M_{\rm NS} R_{\rm NS}} - \Omega \, ({{\rm d}I_{\rm NS}}/{{\rm d}M_{\rm NS}})}{1 - \sqrt{G M_{\rm NS} R_{\rm NS}}\, ({\Omega}/{c^2})} \ .
\end{equation}
The evolution equations \eqref{eq:dM} and \eqref{eq:dOm} are generic for any slowly rotating NS accreting from a Keplerian disk.
In general, however, the accretion and NS's angular-momentum evolution may be more complex.
Such complications can arise if the NS's magnetic field pressure is comparable to the pressure of the radiation and accreting plasma, or if there is feedback from the accretion itself \citep[e.g.,][]{soker_diversity_2019,grichener_common_2019,lopez-camara_disc_2020}.
For a given EoS, we can then find the right-hand sides of the above equations as a function of $M_{\rm NS}$ and $\Omega$, which leads to a closed system of ordinary differential equations, once $\dot{M}_{\rm b}$ is prescribed.
In the CE inspiral context, the baryon mass accretion rate depends on the primary star's envelope structure, which we discuss in the following subsection. 
\subsection{Primary massive-star models, common envelope accretion, inspiral, and ejection}
We evolve single massive stars with the \texttt{MESA} (v12778) stellar-evolution code \citep{paxton_modules_2010,paxton_modules_2013,paxton_modules_2015,paxton_modules_2018,paxton_modules_2019} to obtain their interior structure. 
We consider a total of 6 primary red-giant stars with masses of $M_\star /M_\odot = (12,12,16,16,20,20)$ with respective radii $R_\star/R_\odot = (173,594,322,672,872,1247)$. 
Here, the smaller radii at a given mass corresponds to the RGB base, while the larger radii at a given mass corresponds to the RGB tip. 
For our CE inspiral calculations, we take the envelope structure to be constant in time.
\par
As a NS inspirals in the CE, the envelope plasma supersonically flows past the NS and may be captured within the NS's accretion radius $R_{\rm a} = 2 G M_{\rm NS}/v^2$, where $v$ is the upstream flow velocity, which, in the NS's rest frame is the orbital velocity. 
If the upstream flow is homogeneous, then from Hoyle-Lyttleton (HL) theory \citep{hoyle_effect_1939}, the accretion rate and local drag force obey
\begin{subequations}
\begin{align} \label{eq:dMHL}
\dot{M}_{\rm HL} &= \pi R_{\rm a}^2 \rho v  = \frac{4 \pi \rho G^2 M_{\rm NS}^2}{v^3} \ , \\ \label{eq:FdHL}
F_{\rm d,HL} &= \dot{M}_{\rm HL} v = \pi R_{\rm a}^2 \rho v^2 = \frac{4 \pi \rho G^2 M_{\rm NS}^2}{v^2} \ ,
\end{align}
\end{subequations}
where $\rho$ is the upstream mass density. 
For NS accretion in stellar-envelope environments, the density and temperature may be high enough for neutrino cooling, such that the accretion rate exceeds the Eddington limit \citep{houck_steady_1991}. 
The envelope's local density scale height may be comparable in size to the NS accretion radius, which breaks the symmetry that HL theory assumes and thus requires a treatment of this effect. 
\par
To model the accretion and drag, we use the fitting formulae from \citet[][see their Appendix A]{de_common_2020}.
The accretion and drag coefficients, $C_{\rm a}$ and $C_{\rm d}$ are defined such that the baryonic mass accretion rate and the local drag force are
\begin{subequations} \label{eq:dM_Fd}
\begin{align}
\dot{M}_{\rm b} &= C_{\rm a} \dot{M}_{\rm HL}\,, \quad C_{\rm a} = C_{\rm a} ({\cal M},q, R_{\rm sink})\,, \\ \label{eq:Fddel}
F_{\rm d} &= C_{\rm d} F_{\rm d,HL}\,, \quad  C_{\rm d} = C_{\rm d} ({\cal M},q) \ .
\end{align}
\end{subequations}
These coefficients are both functions of the upstream Mach number ${\cal M}$ and the mass ratio $q$ between the compact object and the enclosed mass within the orbit. 
The accretion coefficient also depends on the sink radius $R_{\rm sink}$, given that the wind-tunnel simulations only resolve the accretion flow up to a sphere with radius $0.05 R_{\rm a}$ surrounding the point-mass accretor. 
Thus, some fraction of matter that flows into the region within $0.05 R_{\rm a}$ ultimately ends up accreting onto the NS.
For each EoS, we use $R_{\rm NS}$ as the sink radius. 
In \autoref{app:coeffs}, we describe in more detail how we compute these accretion and drag coefficients. 
\par
\begin{figure}
\centering
\includegraphics[width=\columnwidth]{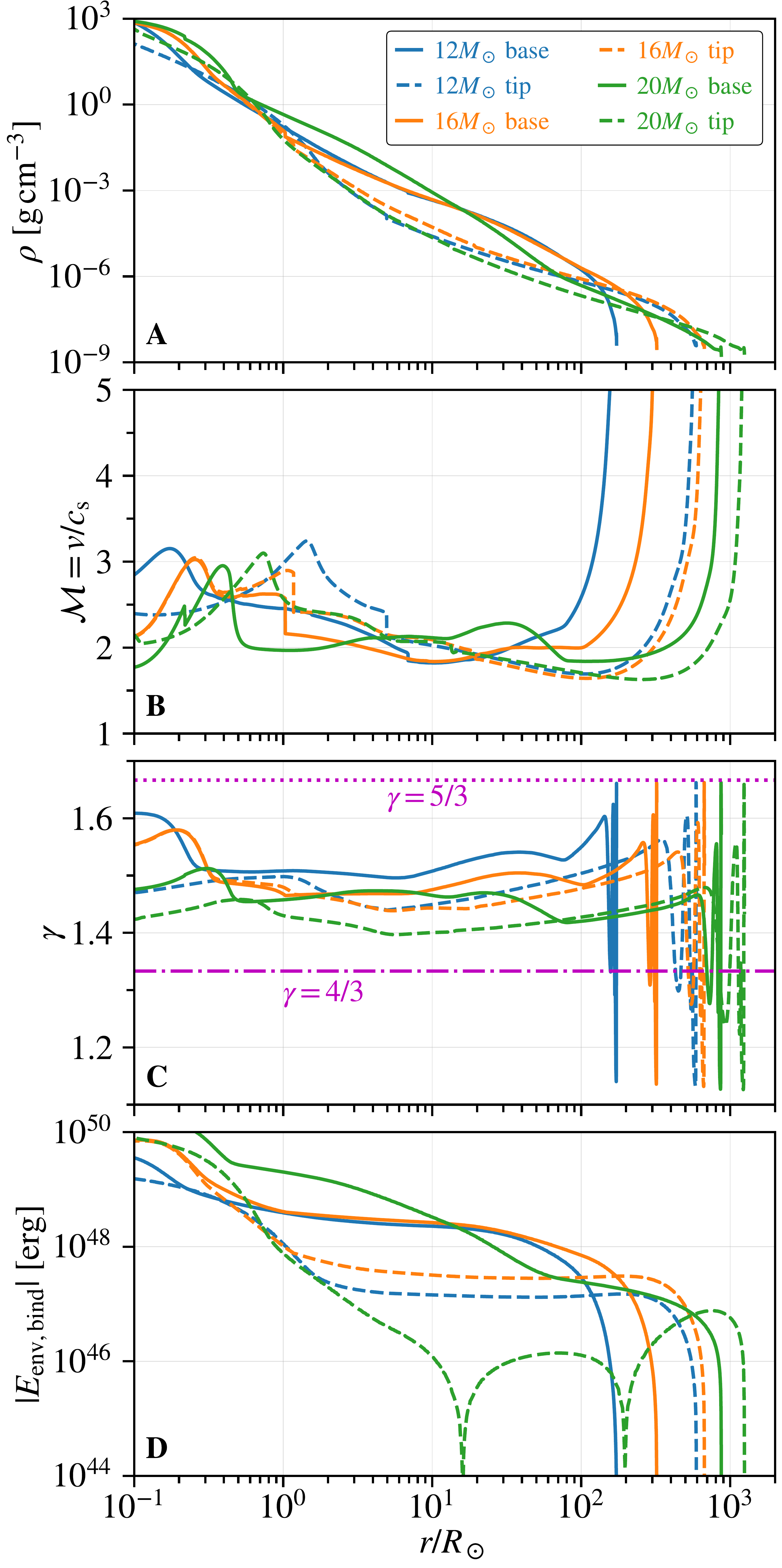}
\caption{\label{fig:mesa} {\bf MESA stellar models.} Panel A: density profiles for primary stellar masses of $M_\star/M_\odot = (12,12,16,16,20,20)$ and respective radii of $R_\star/R_\odot = (173,594,322,672,872,1247)$. 
Masses of $M_\star/M_\odot = (12,16,20)$ correspond to blue, orange, and green colors, respectively. 
Solid and dashed lines correspond to models at the base and tip of the RGB, respectively. 
Panel B: the upstream Mach number ${\cal M} = v/c_{\rm s}$ for each stellar model (formatted in the same manner) for a NS companion with $M_{\rm NS} = 1.4 M_\odot$.
Panel C: the polytropic exponent $\gamma$ for each stellar model. Horizontal magenta lines are shown for $\gamma = 4/3$ and $\gamma = 5/3$ (dot-dashed and dotted, respectively). 
Panel D: envelope binding energy profiles (absolute values) for each stellar model. 
}
\end{figure}

We plot in \autoref{fig:mesa} the stellar profiles of the density, upstream Mach number, polytropic exponent, and envelope binding energy (panels A, B, C, and D, respectively) for the primary masses of $M_\star/M_\odot \in (12,16,20)$ that we consider here.
For a given evolutionary stage and for most of the primary's radii, the $\delta_\rho$ parameter decreases as the primary's mass increases, such that the accretion rate will be greater and result in higher accreted mass. 
Primary stars that are smaller in size will have higher envelope binding energy, which thus requires more energy dissipation during the CE phase in order for successful envelope ejection.
In \S\ref{sec:results}, we quantify how much more NSs accrete when in envelopes with higher binding energies compared to less bound envelopes. 
\par
Given the primary's envelope structure, we can now model the CE inspiral as follows. We approximate the orbital inspiral with Newtonian gravity~\citep{blanchet_gravitational_2014}, given that on the scales of CE evolution, gravity is weak and the orbital velocities are non-relativistic, i.e., $v_{\rm orb}/c \ll 1$. With this in mind, the orbital energy throughout the inspiral is $E = - G M_{\rm NS} m_\star /(2a)$, where $m_\star = m_\star(a) = \int_0^a 4 \pi \rho r^2 \, {\rm d}r$ is the mass enclosed within the NS's separation from the primary's center.
The orbital velocity at any given time obeys $v^2 = G\left[M_{\rm NS} + m_\star(a)\right]/a$, since we consider the inspiral to be quasi-circular. 
The change in the binary orbital energy as the NS inspirals thus obeys
\begin{equation}
{\rm d}E = \frac{\partial E}{\partial m_\star} {\rm d}m_\star + \frac{\partial E}{\partial M_{\rm NS}} {\rm d} M_{\rm NS} + \frac{\partial E}{\partial a} {\rm d}a \ ,
\end{equation}
such that we can then solve for ${\rm d}a/{\rm d}t$
\begin{equation} \label{eq:da}
\frac{{\rm d}a}{{\rm d}t} = a \left( - \frac{\dot E}{E} + \frac{\dot{m}_\star}{m_\star} + \frac{\dot{M}_{\rm NS}}{M_{\rm NS}} \right) = \frac{a}{1 - 4 \pi \rho a^3 / m_\star} \left(- \frac{\dot E}{E} + \frac{\dot{M}_{\rm NS}}{M_{\rm NS}}\right)\ ,
\end{equation}
where $\dot{m}_\star = 4 \pi \rho a^2 \dot{a}$ since we assume a static envelope.
We take the energy decay rate to be the drag luminosity $\dot{E} = - F_{\rm d} (\delta_\rho) v$, which dominates over the gravitational-wave luminosity from the orbital motion, and which can be obtained using both Eqs.~\eqref{eq:FdHL} and \eqref{eq:Fddel}. 
\par
We summarize our integration procedure as follows.
We first precompute the NS properties shown in Eqs.~\eqref{eq:dM} and \eqref{eq:dOm} as well as \autoref{app:phi} for each EoS in our catalog, which are then stored as tables to interpolate from at each timestep of an orbital integration. 
We then explicitly integrate Eqs.~\eqref{eq:dM}, \eqref{eq:dOm}, and \eqref{eq:da} to obtain $M_{\rm NS}$, $\Omega$, and $a$ throughout the CE inspiral. 
The NS properties at each point in the NS's evolution correspond to a Hartle-Thorne NS with gravitational mass $M_{\rm NS}$ and spin $\Omega$ that we obtain from our pre-computed tables. 
Our orbital integrations are carried out for each of our 6 primary stellar models and for each EoS in our catalog, varying the initial NS gravitational mass $M_{\rm NS,0}$ and initial NS spin $\Omega_0$. 
We terminate these orbital integrations when the dissipated orbital energy $\Delta E_{\rm orb} = E(a) - E(a_0)$ is equal to the primary envelope's binding energy $E_{\rm env,bind}$ given by
\begin{equation}
E_{\rm env,bind} = \int_{m_\star(r)}^{m_\star(R_\star)} \left( u - \frac{G M_\star(r)}{r}\right) \, {\rm d}M \ ,
\end{equation}
where $u$ is the stellar fluid's internal energy and where the integration coordinate is the primary's mass coordinate. 
This amounts to assuming a CE efficiency parameter \citep[e.g.,][]{webbink_double_1984} of $\alpha_{\rm CE} = 1$. 

\section{Results} \label{sec:results}
\subsection{NS mass gain and spin-up}
\begin{figure}
\centering
\includegraphics[width=\columnwidth]{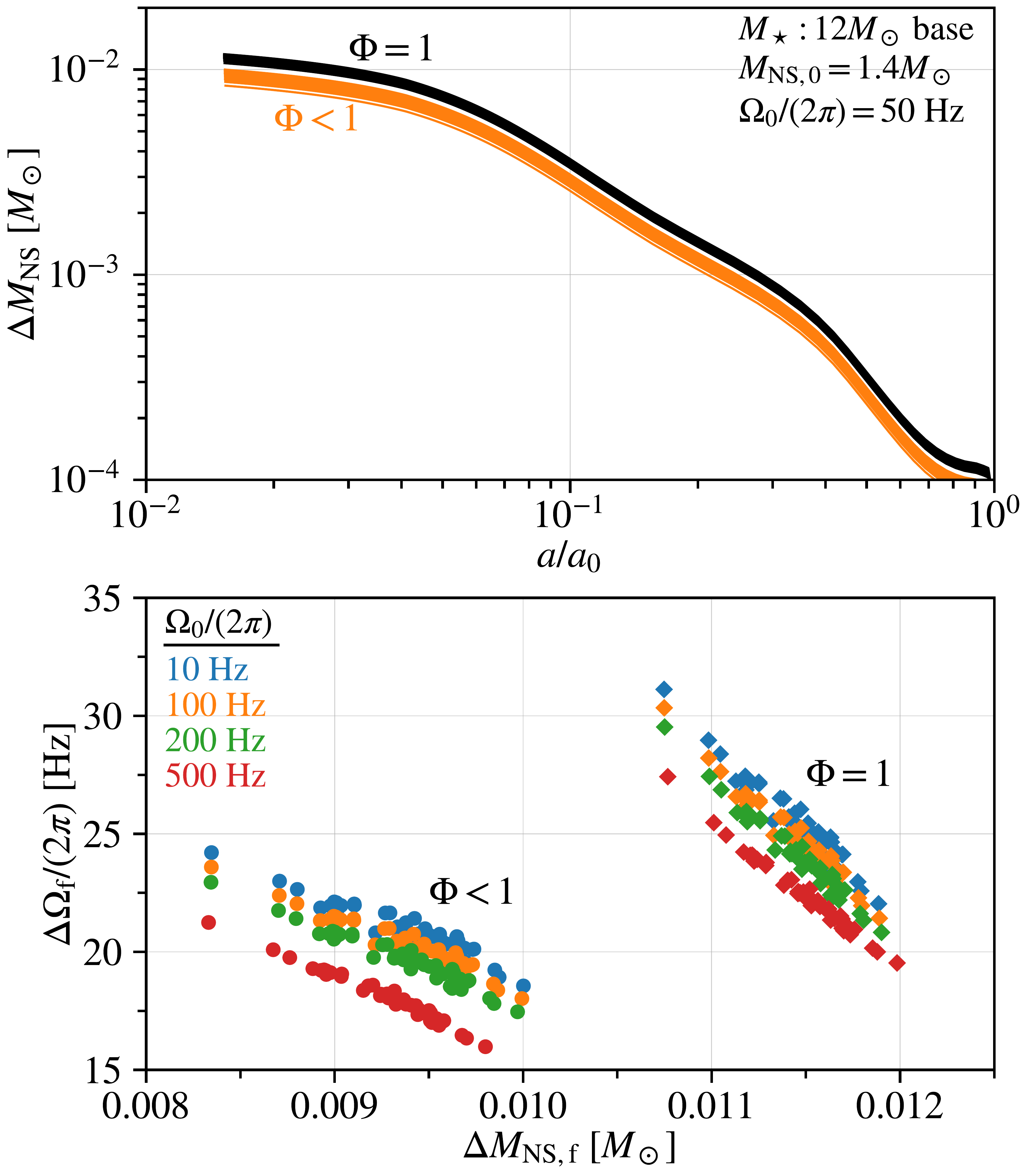}
\caption{\label{fig:evol} {\bf NS mass gain and spin-up.} The initial pre-CE system considered here is a primary star at the base of the RGB with a mass $M_\star = 12 M_\odot$ with a companion NS that has an initial mass $M_{\rm NS,0} = 1.4 M_\odot$ and initial spin $\Omega_0/(2\pi) = 50$ Hz. 
Top panel: gain in gravitational mass vs. orbital separation.
The orange and black curves correspond to $\Phi < 1$ and $\Phi = 1$ (with binding energy vs.~without), respectively, where each curve corresponds to a different EoS. 
Bottom panel: the final spin-up $\Delta\Omega_{\rm f}/(2\pi)$ vs. the final gravitational mass gain $\Delta M_{\rm NS,f}$ for each EoS and the same initial pre-CE parameters, except with a varying initial NS spin. 
The circle and diamond points are for $\Phi = 1$ and $\Phi < 1$, respectively.
The color of each data point corresponds to a different initial NS spin of $\Omega_0/(2\pi) =$ (10, 100, 200, 500) Hz with blue, orange, green, and red, respectively. 
}
\end{figure}

We plot the NS evolution for the often fiducial case of a pre-CE NS mass of $1.4 M_\odot$ and a primary of $12 M_\odot$ in the top panel of \autoref{fig:evol}. 
For this case, the primary is taken to be at the RGB base such that $a_0 = 173 R_\odot$ and we take the initial NS spin to be $\Omega_0/(2\pi) = 50$ Hz.
The black curves corresponds to $\Phi = 1$, i.e., not accounting for NS binding energy. 
Each black and orange curve corresponds to a different EoS in our catalog. 
\par
In all cases, the NS accretes no more than a few percent of its pre-CE mass, due to the suppressed accretion rate from the envelope density gradient. 
The gravitational-mass gain as well as the spin-up further decreases, since some of the accreted baryon mass-energy is converted into binding energy.
In the bottom panel of \autoref{fig:evol}, we plot the final spin-up $\Delta \Omega_{\rm f}/(2\pi) = (\Omega_{\rm f} - \Omega_0)/(2\pi)$ and the final gravitational-mass gain for each EoS model and for varying initial NS spins of $\Omega_0/(2\pi) = (10, 50, 100, 200, 500)$ Hz as blue, orange, green, red, and purple points, respectively. 
Higher initial NS spins increase the NS binding energy, such that less gravitational mass is gained and the spin-up decreases. 
\par
With different EoSs, there is an anti-correlation between the mass gain and spin-up, where an EoS that allows for higher gravitational-mass gain results in a lower spin-up when starting with the same initial NS spin. 
A larger increase in $\Delta M_{\rm NS}$ is a result of a larger $\Phi_{\rm NS}$, i.e., higher baryon mass converted to gravitational mass. 
The gravitochemical potential $\Phi_{\rm NS}$ is proportional to the inertia $I_{\rm NS}$, such that less compact NSs are harder to spin up because they have higher $\Phi_{\rm NS}$ and higher $I_{\rm NS}$. 
Conversely, more compact NSs will gain less gravitational mass and spin-up faster because they have lower $\Phi_{\rm NS}$ and lower $I_{\rm NS}$. 
\subsection{Parameter survey}
\begin{figure*}
\centering
\includegraphics[width=\textwidth]{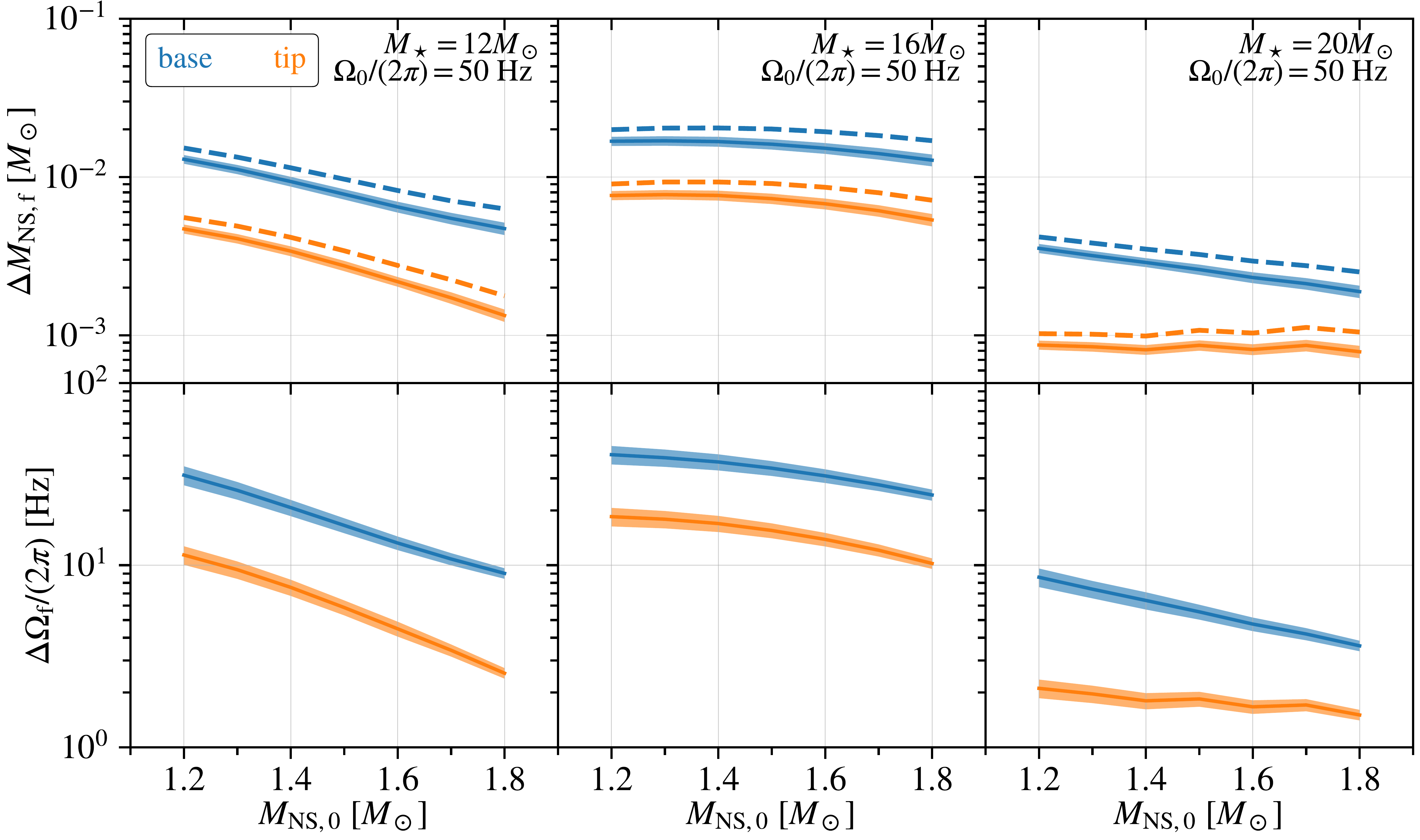}
\caption{\label{fig:gain} {\bf Varying initial NS masses, primary masses, and envelope structures.} 
Top row: the accreted gravitational mass at the end of our orbital integrations for the 6 primary stellar models, initial NS gravitational masses ranging in $M_0/M_\odot \in [1.2,1.8]$ with spacings of $\delta M = 0.1 M_\odot$, and an initial NS spin of 50 Hz. 
The left, middle, and right columns correspond to primary stellar masses of $M_\star/ M_\odot = (12,16,18)$, respectively. 
The top and bottom rows correspond to the accreted mass and the spin-up, respectively. 
The blue and orange curves correspond to primary stellar models at the base and tip of the RGB, respectively.
The dashed lines correspond to $\Phi = 1$, i.e., taking the accreted gravitational mass to be equivalent to the accreted baryonic mass. 
The width of the dashed line encompasses the $\pm2\sigma$ region.
The solid lines with shaded bands correspond to the mean and the $\pm 2\sigma$ deviation, respectively, of our predicted accreted NS masses including binding energy from our catalog of 46 EoSs. 
}
\end{figure*}

Given that the relativistic mass deficit is greater for more massive NSs (see \autoref{fig:phis}), we then vary the pre-CE NS mass. 
We run inspirals for the following set of pre-CE NS masses $M_0/M_\odot \in [1.2,1.8]$ with a step size of $0.1 M_\odot$. 
We plot in the top panels of \autoref{fig:gain} the mean accreted masses when varying the EoS as solid lines with the shaded region corresponding to the $\pm 2\sigma$ deviation. 
The dashed lines correspond to $\Phi = 1$, i.e., taking the accreted gravitational mass to be equivalent to the accreted baryonic mass. 
The width of the dashed line encompasses the $\pm2\sigma$ region.
In the bottom panels of \autoref{fig:gain}, we plot the corresponding spin-up $\Delta\Omega_{\rm f} = (\Omega_{\rm f,0} - \Omega_0)/(2\pi)$.  
\par
An increasing pre-CE NS mass results in a systematically decreasing accreted NS mass across all of our models. 
This is because at constant $\alpha_{\rm CE}$, having a more massive NS results in a larger dissipated orbital energy, such that envelope ejection is achieved at wider separations and such that the accreted baryonic mass is reduced compared to lower-mass NSs.  
It remains to be seen whether or not this trend will hold in global 3D hydrodynamic CE simulations when the initial NS mass is varied. 
Models at the RGB base result in higher accreted mass and spin-up, which is due to the larger envelope binding energy from their smaller sizes as compared to the RGB tip (bottom panel of \autoref{fig:mesa}). 
\par
\begin{figure*}
\centering
\includegraphics[width=\textwidth]{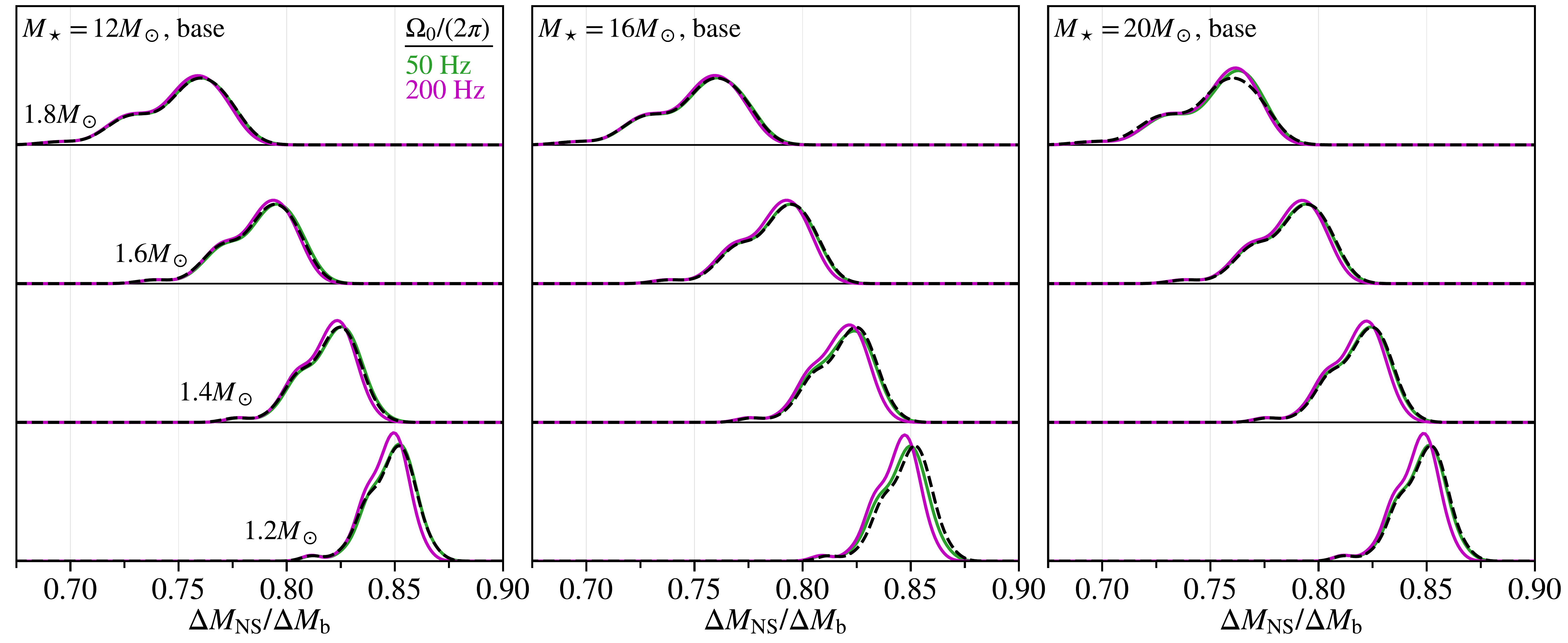}
\caption{\label{fig:dist} {\bf Ratio distributions.} Distributions of the ratio of the accreted gravitational mass to the accreted baryonic mass $\Delta M_{\rm NS}/\Delta M_{\rm b}$. 
These distributions are represented with a kernel density estimator. 
Without accounting for NS binding energy, $\Delta M_{\rm NS}/\Delta M_{\rm b} = 1$. 
The left, middle, and right panels correspond to primary stellar masses of $M_\star/M_\odot = (12,16,18)$ at the RGB base, respectively.
In each panel, each distribution from bottom ascending to top is for initial NS gravitational masses of $M_{\rm NS,0}/M_\odot = (1.2,1.4,1.6,1.8)$, respectively. 
The green and magenta curves correspond to initial NS spins of 50 Hz and 200 Hz, respectively.
The black dashed curve corresponds to the $\Phi_{\rm TOV,0}$ distribution from our EoS catalog, i.e., evaluating \autoref{eq:non} with the initial TOV mass and radius.  
Distributions for the RGB tip case will be similar, though the separation between distributions at the same initial NS mass will be smaller since the accreted baryonic mass for the RGB tip cases is smaller than the RGB base cases (see \autoref{fig:gain}). 
This is quantified in \autoref{app:kl} and \autoref{fig:kldiv}. 
}
\end{figure*}

As previously shown in \autoref{fig:phis}, NSs with higher gravitational mass will convert a larger fraction of the accreted mass into binding energy. 
We plot in \autoref{fig:dist} the distributions of the ratio of the accreted gravitational mass to the accreted baryonic mass from our RGB base models. 
In each panel, each distribution from bottom ascending to top is for initial NS gravitational masses of $M_{\rm NS,0}/M_\odot = (1.2,1.4,1.6.,1.8)$, respectively. 
The green and magenta curves correspond to initial NS spins of 50 Hz and 200 Hz, respectively.
We also plot a black dashed curve that corresponds to using $\Phi_{\rm TOV,0}$ as a fast approximation, i.e., the gravitochemical potential of the non-rotating NS at its initial properties. 
\par
Higher initial spins tend to decrease $\Delta M_{\rm NS}/\Delta M_{\rm b}$, though the model with $M_{\rm NS,0} = 1.8 M_\odot$ and $M_\star = 20 M_\odot$ at the RGB base exhibits opposite behavior, albeit slight. 
Since lower-mass NSs accrete more gravitational mass compared to the higher-mass NSs in our models, the differences between the ratio distributions at various spins is also higher as well. 
Distributions for the RGB tip case will be similar, though the separation between distributions at the same initial NS mass will be smaller since the RGB tip cases resulted in less mass gain (\autoref{fig:gain}). 
\section{Discussion and Conclusions} \label{sec:conclusions}
We have investigated here how NS binding energy affects NS-CE accretion, which plays a role in forming DNSs that merge within a Hubble time. 
We find that the gravitational-mass gain and spin-up is systematically reduced and that this effect is enhanced for higher-mass NSs. 
We also find that more compact NSs will gain less gravitational mass and spin-up faster due to having a lower $\Phi_{\rm NS}$ and a lower $I_{\rm NS}$ compared to less compact NSs. 
The strongest assumption from our model is that the envelope remains static throughout the inspiral. 
Realistically, the envelope is expected to respond and readjust in structure as the NS inspirals deeper toward the primary's core. 
The accretion, which we have focused on in this work, is still expected to be some small fraction of the pre-CE NS mass. 
There will still be density gradients within the envelope that break BHL symmetry and the accreting material still needs to overcome the angular momentum barrier over multiple length scales.
\par
The amount of NS mass gain and spin-up we obtain with this modeling approach may be testable with Galactic DNS observations \citep[e.g.,][]{oslowski_population_2011}. 
For millisecond pulsars, spin-period derivatives corresponding to a spindown timescale of order a Hubble time would be ideal. 
If a phase-transition to quark matter happens in NS interiors, a new branch of stable stars 
with the same masses, but smaller radii relative to their hadronic counterparts can appear \citep[e.g.,][]{Gerlach:1968zz,Kampfer:1981yr,Glendenning:1998ag,Montana:2018bkb}.
These have been called ``twin-stars'' and due to their larger compactness, the effects we 
present here would be further enhanced in comparison to the purely hadronic NSs we studied.
We leave a more systematic investigation of these aspects for future work. 
\par
This work demonstrates that a NS's strong gravity and nuclear microphysics play a role in NS-CE evolution in addition to the macroscopic astrophysics of the envelope. 
Strong gravity and the nuclear EoS thus may affect the population properties of NS binaries and the cosmic double NS merger rate. 
Our results may further inform binary population synthesis models, 1D hydrodynamic CE inspiral coupled to stellar evolution, and global 3D hydrodynamic CE simulations. 
%
\acknowledgments
We thank the anonymous referee for comments and suggestions that led to improvements of this work. 
We thank Cole Miller for insightful discussions and for detailed feedback on our manuscript. 
A.M.H. is supported by the McWilliams Postdoctoral Fellowship. 
This work was also partially supported by the DOE NNSA Stewardship Science Graduate Fellowship under grant No.~DE-NA0003864. 
H.O.S and N.Y. are supported by NASA grants No.~NNX16AB98G, No.~80NSSC17M0041, and No.~80NSSC18K1352 and NSF Award No.~1759615.
P.M.R. acknowledges support by AST 14-13367. 


%



 \software{ \texttt{matplotlib} \citep{hunter_matplotlib:_2007}, \texttt{numpy} \citep{walt_numpy_2011}, \texttt{scipy} \citep{2020SciPy}, \texttt{MESA: v12778} \citep{paxton_modules_2010,paxton_modules_2013,paxton_modules_2015,paxton_modules_2018,paxton_modules_2019} }.

\bibliographystyle{yahapj}
\bibliography{references}
\appendix
\section{The Neutron Star Catalog} \label{app:eos}
\par
We use the same set of 46 EoSs from \cite{silva_astrophysical_2020} for purely hadronic NSs, including ALF2, APR3, APR4, BCPM, BSP, BSR2, BSR2Y, BSk20, BSk21, BSk22, BSk23, BSk24, BSk25, BSk26, DD2, DD2Y, DDHd, DDME2, DDME2Y, ENG, FSUGarnet, G3, GNH3, IOPB, K255, KDE0v1, MPA1, Model1, Rs, SINPA, SK272, SKOp, SKa, SKb, SLY2, SLY230a, SLY4, SLY9, SLy, SkI2, SkI3, SkI4, SkI6, SkMP, WFF1, and WFF2 \citep{read_constraints_2009,kumar_inferring_2019}. 
\section{The Neutron Star Gravitochemical Potential} \label{app:phi}
For a NS with spin parameter $\epsilon = \Omega / \Omega_*$, the gravitochemical potential is defined as~\citep{alecian_effect_2004}
\begin{equation} \label{eq:phins}
\Phi_{\rm NS} = \sqrt{e^\nu (1 + 2 h)} \ ,
\end{equation}
where $\nu$ and $h$ are both metric functions related to the metric tensor via $g_{tt} = -e^\nu (1 + 2 h)$ in the Hartle-Thorne approximation. The metric function $\nu$ is a quantity of ${\cal{O}}(\epsilon^0)$, and thus, it is obtained by solving the TOV equations. The metric correction $h$ is a quantity of ${\cal{O}}(\epsilon^2)$, so we then write it as $h = \epsilon^2 \delta h$, such that the non-rotating limit is recovered as $\epsilon \to 0$ and $\delta h$ remains finite. 

In order to evaluate the gravitochemical potential $\Phi_{\rm NS}$, we need to solve for the function $h$, which therefore requires that we solve the Einstein equations at second-order in the small rotation expansion~\citep{hartle_slowly_1967}. Performing a Legendre decomposition, we can write
\begin{equation} \label{eq:hr}
\delta h(r) = \delta h_0(r) + \delta h_2(r) P_2(\cos \theta) \ ,
\end{equation}
where $\theta$ is the polar angle from the equator, and $P_2$ is the second-order Legendre polynomial. Matching the interior and the exterior solutions at the NS surface allows us to find an exact solution for $\delta h_0(r)$ at the NS surface, namely
\begin{equation}
\delta h_0(R_{\rm TOV}) = - \frac{\delta M}{R_{\rm TOV} - 2M_{\rm TOV}} + \frac{\delta J^2}{R_{\rm TOV}^3 (R_{\rm TOV}-2M_{\rm TOV})} \ .
\end{equation}
Here, $\delta J$ is the NS angular momentum at the Keplerian angular spin frequency. 
The function $\delta h_2(r)$ generally obeys $\delta h_2 \ll \delta h_0$, such that when this function is scaled by $\epsilon^2$, which, for this work obeys $\epsilon^2 \ll 1$, the contribution from the $\epsilon^2 \delta h_2$ component to \autoref{eq:phins} is effectively negligible.
We thus take $\delta h(R_{\rm NS}) \approx \delta h_0(R_{\rm NS})$, such that
\begin{equation}
 h (R_{\rm NS}) \approx  \epsilon^2 \delta h_0 (R_{\rm TOV}) = - \frac{\epsilon^2 \delta M}{R_{\rm TOV} - 2 M_{\rm TOV}} + \frac{\epsilon^2 \delta J^2}{R_{\rm TOV}^3 (R_{\rm TOV} - 2 M_{\rm TOV})} \ .
\end{equation}
\section{Accretion and Drag Coefficients} \label{app:coeffs}
\cite{de_common_2020} present fitting formulae for the accretion rate and drag within a non-relativistic background plasma.
They consider two polytropic exponents of $\gamma = 4/3$ and $\gamma = 5/3$, where the coefficients for each fitting formula are given in their Tables A1 and A2. 
Given that our stellar models for the massive primaries have polytropic exponents that predominantly obey $4/3 \leqslant \gamma \leqslant 5/3$ (see \autoref{fig:mesa}), we compute the accretion and drag coefficients by weighting both the $C_{\rm ad,4/3}$ and $C_{\rm ad,5/3}$ formulae as
\begin{subequations} \label{eq:cad}
\begin{align}
C_{\rm a} &= \xi \left(w_{4/3} C_{\rm a,4/3} + w_{5/3} C_{\rm a,5/3} \right) \ , \\
C_{\rm d} &= w_{4/3} C_{\rm d,4/3} + w_{5/3} C_{\rm d,5/3} \ ,
\end{align}
\end{subequations}
where
\begin{subequations}
\begin{align}
w_{4/3} = 1 - 3(\gamma - 4/3) \ , \\
w_{5/3} = 1-3(5/3-\gamma) \ ,
\end{align}
\end{subequations}
and where $\xi$ is defined as
\begin{equation}
\xi \equiv (R_{\rm NS}/0.05 R_{\rm a})^{0.33} \ .
\end{equation}
For $\gamma < 4/3$, we use $C_{\rm ad,4/3}$. 
The factor $\xi$ approximates the fraction of matter flowing into the sink radius that ultimately accretes onto the NS. 
Given that these wind-tunnel models do not resolve the flow past a sink radius $R_{\rm sink} = 0.05 R_{\rm a}$, the matter falling into this sink volume is likely to be an upper estimate of the NS's accreted baryons. 
\cite{de_common_2020} estimate how the accretion rate depends on the sink radius and fit a power-law dependence $\dot {M} \propto \left(R_{\rm sink}/ R_{\rm a}\right)^{\alpha_{\dot{M}}}$, where $\alpha_{\dot{M}} \approx 0.33$ with a scatter of order tens of percent.
\section{Kullback-Leibler Divergence} \label{app:kl}
For a given NS-CE system evolution with an initial primary star with mass $M_\star$ and radius $R_\star$ and an initial NS with mass $M_{\rm NS,0}$ and spin $\Omega_0$, we define $p$ as distribution of $\Delta M_{\rm NS}/\Delta M_{\rm b}$ predicted from our EoS catalog and semi-analytic models.
We also define $q$ as the distribution of $\Phi_{\rm TOV,0}$ from our EoS catalog, i.e., evaluating \autoref{eq:non} at the initial NS parameters. 
Given these two distributions, we can compute the Kullback-Leibler divergence, given by 
\begin{equation}
{\cal D} (p||q) = \int p(x) \ln \left(\frac{p(x)}{q(x)}\right) \ {\rm d} x \ ,
\end{equation}
where $x \equiv \Delta M_{\rm NS}/\Delta M_{\rm b}$. 
The distributions $p$ and $q$ are approximated as a kernel-density estimate of the samples for each model. 
One can interpret the KL divergence between $p$ and $q$ as the information loss when using $q$ to approximate $p$.
Conversely, it can be interpreted as the information gained by using $p$ in place of $q$. 
\par
\begin{figure*}
\centering
\includegraphics[width=\textwidth]{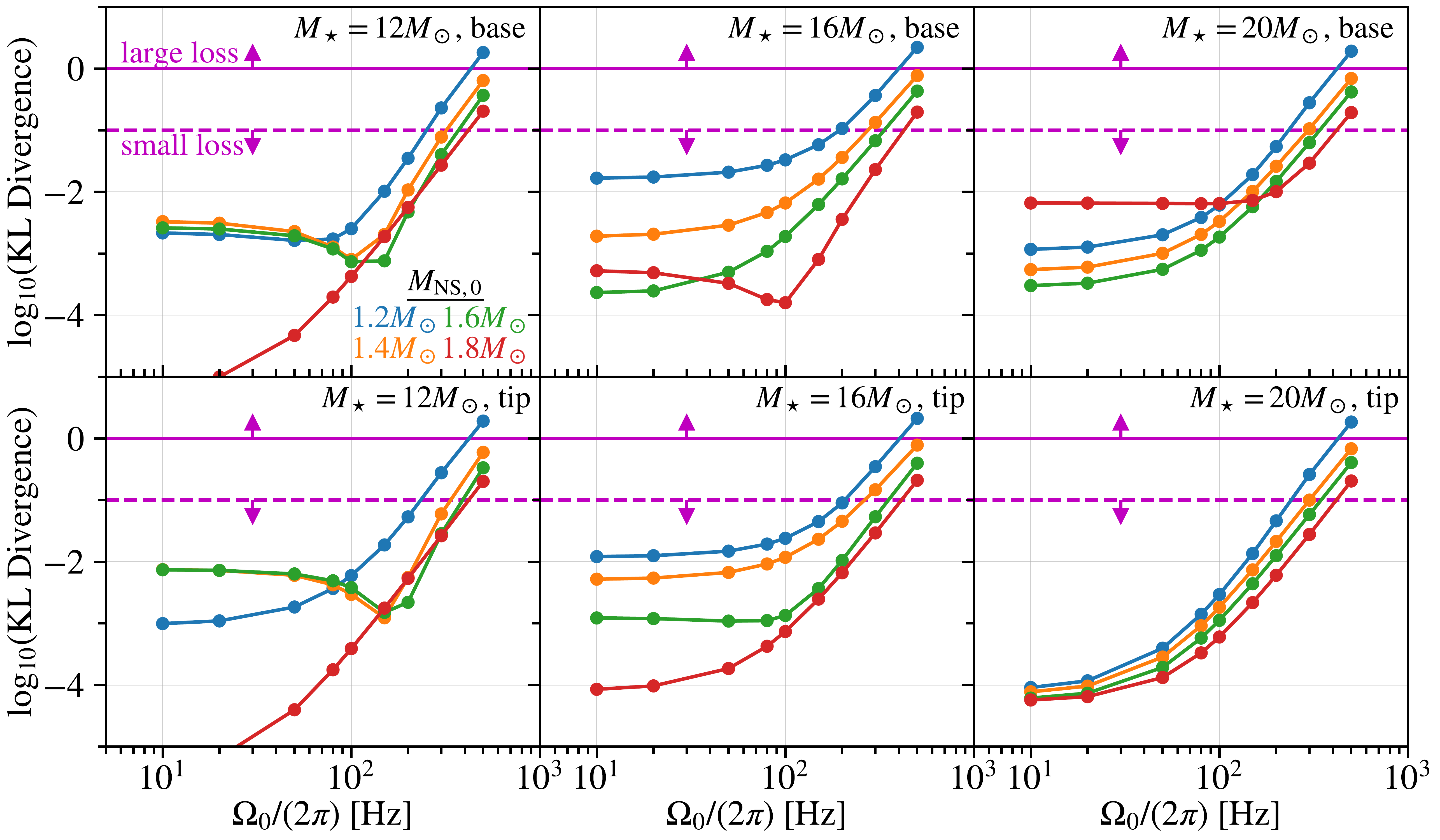}
\caption{\label{fig:kldiv} {\bf KL divergence vs. initial NS spin.} 
The KL divergence between $\Delta M_{\rm NS}/\Delta M_{\rm b}$ and $\Phi_{\rm TOV,0}$ evaluated at varying initial NS spins of $\Omega_0/(2\pi) = (10, 20, 50, 80, 100, 150, 200, 300, 500)$ Hz.
The top and bottom rows are for stellar models at the RGB base and tip, respectively, with each column for primary stellar masses of $M_\star/ M_\odot = (12,16,18)$, respectively. 
The blue, orange, green, and red curves correspond to initial NS gravitational masses of $M_{\rm NS,0}/M_\odot = (1.2, 1.4, 1.6, 1.8)$. 
For KL divergences $\lesssim 0.1$, the information loss is considered to be small, while KL divergences $\gtrsim 1$ corresponds to a large information loss. 
}
\end{figure*}

Directly using $\Phi_{\rm TOV,0}$ as a fast approximation in other models such as population synthesis or as a subgrid prescription for global 3D hydrodynamic simulations may be acceptable as long as $\Delta M_{\rm NS}/ M_{\rm NS,0} \lesssim 1\%$ and if the initial NS spin is low enough. 
To quantify the information loss from this approximation, we compute the KL divergences \citep[][\autoref{app:kl}]{kullback_information_1951} between our semi-analytic inspiral models and using \autoref{eq:non} at the initial NS properties over a range of initial NS spins: $\Omega_0/(2\pi) = (10, 20, 50, 80, 100, 150, 200, 300, 500)$ Hz. 
We plot the KL divergences for each of our models in \autoref{fig:kldiv}.
For initial NS spins of $\lesssim 200$ Hz, the KL divergence is $\lesssim 0.1$, corresponding to a small information loss and thus $\Phi_{\rm TOV,0}$ being a reasonable approximation if used in other models. 
\end{document}